\documentclass[12pt]{article}
\usepackage[utf8]{inputenc}
\usepackage{amsfonts,amsmath,amssymb,mathtools}
\usepackage[left=2cm,right=2cm,top=2cm,bottom=2cm,bindingoffset=0cm]{geometry}
\usepackage{graphicx}
\usepackage{cite}
\usepackage{authblk}
\usepackage{xcolor}
\usepackage{hyperref}

\newcommand\bem{\begin{pmatrix}}
\newcommand\eem{\end{pmatrix}}
\newcommand\beq{\begin{equation}}
\newcommand\eeq{\end{equation}}
\newcommand\beqs{\begin{equation*}}
\newcommand\eeqs{\end{equation*}}

\newcommand{\pd}{\partial}

\newcommand{\mO}{\mathcal{O}}
\newcommand{\mN}{\mathcal{N}}
\newcommand{\const}{\text{const}}

\numberwithin{equation}{section}

\hypersetup{
    colorlinks=true,
    linkcolor=black,
    citecolor=blue,   
    urlcolor=blue,
}

\title{\bf Particle creation in nonstationary large N quantum mechanics}
\author[1,2]{Dmitrii~A.~Trunin\thanks{\href{mailto:dmitriy.trunin@phystech.edu}{dmitriy.trunin@phystech.edu}}}
\affil[1]{Moscow Institute of Physics and Technology, 141700, Institutskiy pereulok, 9, Dolgoprudny, Russia}
\affil[2]{Institute for Theoretical and Experimental Physics, 117218, B. Cheremushkinskaya, 25, Moscow, Russia}
\date{\today}

\begin{document}

\maketitle

\begin{abstract}
We consider an analog of particle production in a quartic $O(N)$ quantum oscillator with time-dependent frequency, which is a toy model of particle production in the dynamical Casimir effect and de Sitter space. We calculate exact quantum averages, Keldysh propagator, and particle number using two different methods. First, we employ a kind of rotating wave approximation to estimate these quantities for small deviations from stationarity. Second, we extend these results to arbitrarily large deviations using the Schwinger-Keldysh diagrammatic technique. We show that in strongly nonstationary situations, including resonant oscillations, loop corrections to the tree-level expressions effectively result in an additional degree of freedom, $N \to N + \frac{3}{2}$, which modifies the average number and energy of created particles.
\end{abstract}

\newpage


\section{Introduction}
\label{sec:intro}

Particle production in nonstationary background fields is an old and fruitful topic that covers many intriguing phenomena. The first example of such a phenomenon --- the spontaneous creation of electron-positron pairs in strong electric fields --- was discovered in 1951 by J.~S.~Schwinger~\cite{Schwinger:1951}. Later it was also supplemented by famous Hawking~\cite{Hawking, Hawking:1974, Hawking:1976}, Unruh~\cite{Unruh, Fulling, Davies:1974}, and dynamical Casimir~\cite{Moore, Davies:1976, Davies:1977, DeWitt} effects. On one hand, all these effects have no classical analogs and reflect the most fundamental features of the quantum field theory. On the other hand, they may hint at a more general theory~\cite{Preskill, Harlow, Almheiri}. A comprehensive review of these celebrated effects is presented in the textbooks~\cite{Birrell, Fulling:1989, Grib}.

A common approach to the particle creation\footnote{We emphasize that in general free Hamiltonian may be nondiagonalizable in the asymptotic future or past~\cite{Polyakov:Eternity, Krotov, Polyakov, Akhmedov:dS}. In this case the notion of particle is meaningless, and the discussed semiclassical method does not work. Instead, one needs to calculate correlation functions whose meaning is always transparent.} in an external background relies on the semiclassical (tree-level) approximation that neglects interactions between quantum fields~\cite{Birrell}. Let us briefly review this approach. Consider a free quantum field $\phi$ with the following mode decomposition:
\beq \phi(t,\mathbf{x}) = \begin{cases} \sum_n \left[ a_n^\text{in} f_n^\text{in}(t,\mathbf{x}) + H.c. \right], \quad & \text{as} \quad t \rightarrow -\infty, \\ \sum_n \left[ a_n^\text{out} f_n^\text{out}(t,\mathbf{x}) + H.c. \right], \quad & \text{as} \quad t \rightarrow +\infty, \end{cases} \eeq
where functions $f_n^\text{in}$, $f_n^\text{out}$ solve the free equations of motion, approximately diagonalize the free Hamiltonian in the corresponding limits and form a complete basis with respect to the appropriate inner product (e.g. Klein-Gordon inner product in the case of a free scalar field). Usually functions $f_n^\text{in}$ and $f_n^\text{out}$ are referred to as in- and out-modes, respectively. In- and out- annihilation and creation operators, $a_n^\text{in}$, $(a_n^\text{in})^\dag$ and $a_n^\text{out}$, $(a_n^\text{out})^\dag$, satisfy the standard commutation relations. For simplicity, we also assume that in the asymptotic past the field was in the pure vacuum state $| in \rangle$ defined with respect to the initial annihilation operators, $a_n^\text{in} | in \rangle = 0$ for all $n$.

Let us calculate the number of created out-particles ($f_n^\text{out}$ modes). In the Heisenberg picture it is given by the following expression:
\beq \mN_n^\text{free} = \langle in | (a_n^\text{out})^\dag a_n^\text{out} | in \rangle \quad \text{(no sum)}. \eeq
In a nonstationary situation in- and out- modes and creation operators are related by a generalized Bogoliubov (or canonical) transformation:
\beq \label{eq:Bogoliubov} \begin{aligned}
f_n^\text{out} &= \sum_k \left[ \alpha_{nk}^* f_k^\text{in} - \beta_{nk} (f_k^\text{in})^* \right], \\
a_n^\text{out} &= \sum_k \left[ \alpha_{nk} a_k^\text{in} + \beta_{nk}^* (a_k^\text{in})^\dag \right],
\end{aligned} \eeq
with nonzero Bogoliubov coefficients $\alpha_{nk}$ and $\beta_{nk}$. This implies the following identity:
\beq \label{eq:N-free}
\mN_n^\text{free} = \sum_{k,l} \langle in | \left[ \alpha_{nk}^* (a_k^\text{in})^\dag + \beta_{nk} a_k^\text{in} \right] \left[ \alpha_{nl} a_l^\text{in} + \beta_{nl}^* (a_l^\text{in})^\dag \right] | in \rangle = \sum_k |\beta_{nk}|^2. \eeq
Thus on the tree level particle creation is only due to the amplification of vacuum fluctuations.

Now let us turn on interactions:
\beq \mN_n(t) = \big\langle in \big| U^\dag(t, t_0) \big(a_n^\text{out}(t)\big)^\dag a_n^\text{out}(t) U(t, t_0) \big| in \big\rangle, \eeq
where $U(t, t_0)$ denotes the evolution operator in the interaction picture. Note that we restored the initial and final moments because in the interacting case limits $t_0 \to -\infty$ and $t \to \infty$ should be taken with caution~\cite{Polyakov:Eternity, Krotov, Polyakov, Akhmedov:dS}. Moreover, interactions between the fields may generate nontrivial quantum averages:
\begin{align}
n_{kl}(t) &= \langle in | U^\dag(t, t_0) (a_k^\text{in})^\dag a_l^\text{in} U(t, t_0) | in \rangle \neq 0, \label{eq:n} \\
\kappa_{kl}(t) &= \langle in | U^\dag(t, t_0) a_k^\text{in} a_l^\text{in} U(t, t_0) | in \rangle \neq 0. \label{eq:k}
\end{align}
Quantities $n_{kl}$ and $\kappa_{kl}$ are usually referred to as level population and anomalous quantum average (or correlated pair density), respectively.

Therefore, in the interacting theory identity~\eqref{eq:N-free} should be modified as follows:
\beq \label{eq:N-exact}
\mN_n = \sum_k |\beta_{nk}|^2 + \sum_{k,l} \left( \alpha_{nk} \alpha_{nl}^* + \beta_{nk} \beta_{nl}^* \right) n_{kl} + \sum_{k,l} \beta_{nk} \alpha_{nl} \kappa_{kl} + \sum_{k,l} \beta_{nl}^* \alpha_{nk}^* \kappa_{lk}^*. \eeq
Common wisdom states that for small coupling constants quantum loop corrections to the averages~\eqref{eq:n} and~\eqref{eq:k} are small and hence can be neglected. However, recently this assumption has been shown to be wrong: even if the coupling constant goes to zero, $\lambda \to 0$, for large evolution times, $t \rightarrow \infty$, loop corrections to $n_{kl}$ and $\kappa_{kl}$ grow secularly and remain finite, $n_{kl}^{(n)} \sim (\lambda t)^{a_n}$ and $\kappa_{kl}^{(n)} \sim (\lambda t)^{b_n}$, with some constants $a_n > 0$ or $b_n > 0$ for every $n$-loop contribution. In particular, such a secular growth was observed in the expanding Universe~\cite{Krotov, Polyakov, Akhmedov:dS, Popov, Akhmedov-1, Burda, Akhmedov-2, Bascone, Pavlenko, Moschella}, the dynamical Casimir effect~\cite{Alexeev, Akopyan}, and nonstationary quantum mechanics~\cite{Trunin-1}, as well as in strong electric~\cite{Musaev, Akhmedov:Et, Akhmedov:Ex}, scalar~\cite{Diatlyk-1, Trunin-2, Diatlyk-2}, and gravitational~\cite{Akhmedov:H} fields. Moreover, in some cases the exact resummed $n_{kl}$ explodes and significantly surpasses the tree-level expression~\cite{Akhmedov:dS, Akhmedov-1, Akhmedov-2, Burda, Popov}.

Thus we need to calculate quantum loop corrections to the averages~\eqref{eq:n} and~\eqref{eq:k} to estimate the correct number of created particles in a nonstationary interacting theory. Unfortunately, this cannot be done perturbatively because higher loop corrections to $n_{kl}$ and $\kappa_{kl}$ are not suppressed when $\lambda$ is finite. The only known way to estimate these quantities is to solve the system of Dyson-Schwinger equations for the propagators and vertices. We emphasize that this system should be deduced using the Schwinger-Keldysh diagrammatic technique~\cite{Schwinger, Keldysh, Kamenev, Rammer, Landau:vol10, Arseev} due to the nonstationarity of the theory. 

In general, the system of the Dyson-Schwinger equations is very complex and cannot be explicitly solved. Fortunately, in some models loop corrections to vertices, retarded/advanced propagators and $\kappa_{kl}$ are suppressed by higher powers of $\lambda$, so in the limit $\lambda \to 0$, $t \to \infty$ the full system is reduced to a single equation on the level population $n_{kl}$. In such models Dyson-Schwinger equations reproduce an equation of kinetic type with additional terms which describe creation and annihilation of particles by external sources. The solution to this --- relatively simple --- equation implies the exact $n_{kl}$ and allows one to estimate $\mN_n$. The notable examples of such ``kinetic'' systems are heavy fields in de Sitter space~\cite{Akhmedov:dS, Akhmedov-1, Akhmedov-2, Burda, Popov, Bascone} and scalar quantum electrodynamics~\cite{Akhmedov:Et, Akhmedov:Ex}.

Of course, there is also a set of models where kinetic approximation is not applicable and loop corrections to the vertices are not suppressed. It includes such important systems as light fields in de Sitter space~\cite{Popov, Pavlenko, Moschella} and the dynamical Casimir effect~\cite{Alexeev, Akopyan}. To the best of our knowledge, there is no systematic approach to the particle creation in such models. The only case where quantities $n_{kl}$ and $\kappa_{kl}$ were analytically estimated is the large~$N$ limit of the $O(N)$ light scalar field with quartic self-interaction and the Bunch-Davies initial state in de Sitter space~\cite{Serreau-1, Serreau-2, Serreau-3, Nacir}.

In this paper we develop a systematic approach to the calculation of $n_{kl}$ and $\kappa_{kl}$ in ``nonkinetic'' systems. This approach is based on the idea that rapidly oscillating parts of the effective Hamiltonian are negligible in the limit $\lambda \to 0$, $t \to \infty$. This approximation resembles the rotating wave approximation from the quantum optics~\cite{Walls, Irish, Law}. The other key approximation is the large~$N$ limit that allows us to single out a particular set of diagrams.

We illustrate our approach on a simple nonstationary large~$N$ system --- a quantum anharmonic oscillator with a quartic $O(N)$ interaction term and time-dependent frequency. On the one hand, a tree-level version of this model is a famous toy model of the dynamical Casimir effect~\cite{Law, Dodonov:1993, Dodonov:1996, Dodonov:2020} and squeezed states generation~\cite{Dodonov:1982, Dodonov:1990, Dodonov:1995} (also see~\cite{Deutsch, Tanas, Recaimer, Sousa} for examples of similar toy models). Therefore, it is important to check how nonlinearities affect the predictions made for the tree-level model. On the other hand, the properties of this simple model are similar to those of higher-dimensional nonstationary quantum field theories with a ``nonkinetic'' behavior of loop corrections~\cite{Popov, Pavlenko, Moschella, Alexeev, Akopyan}. Due to this reason, we believe that our method of loop summation will help one understand nonstationary phenomena in these complex theories.

We emphasize that in this model the expectation value of the evolved free Hamiltonian at the future infinity\footnote{The full Hamiltonian, $\bar{H}_\text{full} = \bar{H} + \bar{H}_\text{int}$, also contains the interaction term $\bar{H}_\text{int}(t) \equiv \frac{\lambda}{4 N} \langle in | \left( \phi_i(t) \phi_i(t)\right)^2 | in \rangle$. However, in the limit $\lambda \to 0$, which we discuss in this paper, the contribution of the interaction term is negligible.}, $\bar{H} \equiv \langle in | U^\dag(t,t_0) H_\text{free} U(t,t_0) | in \rangle$, is expressed through the $\mN = \sum_{n=1}^N \mN_n$:
\beq \bar{H}(t) = \frac{1}{2} \omega_+ N + \omega_+ \mN, \quad \text{as} \quad t \to +\infty \quad \text{and} \quad \lambda \to 0, \eeq
where $\omega_+$ is the frequency of the oscillator in the asymptotic future and $\mN_n$ is defined by~\eqref{eq:N-exact}. This confirms the interpretation of $\mN$ as the total number of created out-particles in an interacting theory. The details on the derivation of this expression are presented in appendix~\ref{sec:N}.

This paper is organized as follows. In section~\ref{sec:QM} we introduce the model and discuss the field quantization on a nonstationary background. In section~\ref{sec:H} we derive a simple effective Hamiltonian of the model and calculate exact $n_{kl}$ and $\kappa_{kl}$ for small deviations from stationarity. In section~\ref{sec:SK} we generalize these calculations to arbitrarily large deviations from stationarity using the Schwinger-Keldysh diagrammatic technique. Finally, we discuss the results and conclude in section~\ref{sec:discussion}. We also explain the physical meaning of $\mN$ in appendix~\ref{sec:N} and discuss a finite~$N$ version of our model in appendix~\ref{sec:small-b}.

\section{Field quantization}
\label{sec:QM}

Consider $N$ copies of a quantum oscillator with time-dependent frequency coupled through an $O(N)$ quartic interaction term:
\beq \label{eq:L}
\mathcal{L} = \frac{1}{2} \dot{\phi}_i \dot{\phi}_i - \frac{\omega^2(t)}{2} \phi_i \phi_i - \frac{\lambda}{4 N} (\phi_i \phi_i)^2, \eeq
where we assume the summation over the repeated indices and introduce the `t Hooft coupling $\lambda = g N$. We will consider asymptotically static situations, i.e., $\omega(t) \to \omega_\pm$ as $t \to \pm \infty$, and assume that the self-interaction term is turned on adiabatically after the time $t_0$.

Although this problem is purely quantum mechanical, we will treat it as a $(0+1)$-dimensional quantum field theory. Similarly to higher-dimensional theories, we introduce the mode decomposition for the free scalar field:
\beq \label{eq:phi}
\phi_i(t) = a_i f(t) + a_i^\dag f^*(t), \eeq
where $a_i^\dag$, $a_i$ are the creation and annihilation operators with the standard commutation relation, $[a_i, a_j^\dag] = \delta_{ij}$, and the mode function $f(t)$ solves the classical free equation of motion:
\beq \label{eq:osc-f} \ddot{f}(t) + \omega^2
(t) f(t) = 0. \eeq
In the asymptotic past and future, oscillation frequency is approximately constant, so the solution of the equation~\eqref{eq:osc-f} is given by the sum of two oscillating exponents\footnote{These are the so-called in-modes that were mentioned in the introduction. However, in what follows we will suppress the index ``in'' for brevity: $f(t) = f^\text{in}(t)$.}:
\beq \label{eq:qm-modes}
f(t) = \begin{cases} \frac{1}{\sqrt{2 \omega_-}} e^{- i \omega_- t}, \quad &\text{as} \quad t \to -\infty, \\ \frac{\alpha}{\sqrt{2 \omega_+}} e^{- i \omega_+ t} + \frac{\beta}{\sqrt{2 \omega_+}} e^{i \omega_+ t}, \quad &\text{as} \quad t \to +\infty, \end{cases} \eeq
where complex numbers $\alpha$ and $\beta$ satisfy the relation $|\alpha|^2 - |\beta|^2 = 1$ as a consequence of the canonical commutation relation $[ \phi_i, \pi_i] = [\phi_i, \dot{\phi}_i] = i$. Note that these coefficients coincide with the Bogoliubov coefficients from the transformation~\eqref{eq:Bogoliubov} if we choose $f^\text{out}(t) = \frac{1}{\sqrt{2 \omega_+}} e^{-i \omega_+ t}$ as $t \to +\infty$. Also note that modes~\eqref{eq:qm-modes} diagonalize the free Hamiltonian at the asymptotic past:
\beq H_\text{free} = \frac{1}{2} \dot{\phi}_i \dot{\phi}_i + \frac{\omega^2(t)}{2} \phi_i \phi_i \approx \omega_- \left( a_i^\dag a_i + \frac{N}{2} \right), \quad \text{as} \quad t \to -\infty. \eeq
For simplicity in the remainder of this paper we assume that the initial state of the field coincides with the ground state of this Hamiltonian at past infinity, $| in \rangle = | 0 \rangle$, $a_i | 0 \rangle = 0$ for all $i$.

Coefficients $\alpha$ and $\beta$ can be unambiguously restored from the function $\omega(t)$, although for arbitrary functions this task can be very difficult. Nevertheless, it significantly simplifies if the variations of the frequency are small, i.e., $\omega(t) = \omega + \delta \omega(t)$ with $\omega = \const$ and $\delta \omega(t) \ll \omega$. On the one hand, in the nonresonant case $\beta$ can be approximated as follows:
\beq \label{eq:b-small}
|\beta|^2 \approx \frac{|\beta|^2}{|\alpha|^2} \approx \left| \int_{-\infty}^\infty \delta \omega(t) e^{-2 i \omega t} dt  \right|^2 \ll 1, \eeq
where we have used an approximate expression for the reflection coefficient~\cite{Migdal}. We emphasize that reflected waves are almost negligible for such variations of frequency; this behavior illustrates the well-known adiabatic theorem~\cite{Trunin-1, Born}. On the other hand, in the resonant case, e.g., $\omega(t) = \omega \left[ 1 + 2 \gamma \cos(2 \omega t) \right]$, $\gamma \ll 1$, both coefficients exponentially grow with time~\cite{Dodonov:1982, Dodonov:1996}:
\beq \label{eq:b-large}
\alpha = \cosh\left( \omega \gamma t_R \right), \quad \beta = -i \sinh\left( \omega \gamma t_R \right), \eeq
where $t_R$ is the duration of the resonant oscillations. These coefficients straightforwardly follow from the equation~\eqref{eq:osc-f} after the substitution of the ansatz $f(t) = \frac{1}{\sqrt{2 \omega}} \left[ \alpha(t) e^{-i \omega t} + \beta(t) e^{i \omega t} \right]$ and averaging over the fast oscillations.

In what follows we will use expressions~\eqref{eq:b-small} and~\eqref{eq:b-large} to estimate the number of created ``particles'' (or, more accurately, excitations) in the infinite future. In other words, we will consider times where $\omega$, $\alpha$ and $\beta$ are approximately constant.

\section{Effective Hamiltonian}
\label{sec:H}

Now let us employ the mode function~\eqref{eq:qm-modes} to calculate the free:
\beq H_\text{free} = \left( |\alpha|^2 + |\beta|^2 \right) \omega_+ \left(a_i^\dag a_i + \frac{N}{2} \right) + \alpha \beta \omega_+ a_i a_i + \alpha^* \beta^* \omega_+ a_i^\dag a_i^\dag, \eeq
and interacting Hamiltonians:
\begin{align}
& \begin{aligned} \label{eq:qm-Hint}
H_\text{int} &\approx \frac{\lambda}{16 N \omega_+^2} \left( |\alpha|^4 + 4 |\alpha|^2 |\beta|^2 + |\beta|^4 \right) \left( a_i^\dag a_i^\dag a_j a_j + 2 a_i^\dag a_j^\dag a_i a_j \right) + \\ &+ \frac{3 \lambda \alpha \beta}{4 N \omega_+^2} \left( |\alpha|^2 + |\beta|^2 \right) a_i^\dag a_i a_j a_j + \frac{3 \lambda \alpha^2 \beta^2}{8 N \omega_+^2} a_i a_i a_j a_j + H.c. + \delta H_\text{free},
\end{aligned} \\
& \begin{aligned} \label{eq:qm-H-renorm}
\delta H_\text{free} \approx \frac{3 \lambda (N + 2) }{8 N \omega_+^2} \alpha \beta \left( |\alpha|^2 + |\beta|^2 \right) a_i a_i + \frac{\lambda (N+2)}{8 N \omega_+^2} \left( |\alpha|^4 + 4 |\alpha|^2 |\beta|^2 + |\beta|^4 \right) \left(a_i^\dag a_i + \frac{N}{2} \right) + H.c., \end{aligned}
\end{align}
in the interaction picture at the future infinity ($t \to +\infty$). Here we neglect oscillating terms that give suppressed contributions to the correlation functions in the limit in question. In other words, we keep only the terms that give the leading contribution to the operator $\int_{t_0}^t H_\text{int}(t') dt'$ in the limit $\lambda \to 0$, $t \to \infty$, $\lambda t = \const$\footnote{This constant has dimensionality of $\text{length}^{-2}$, i.e., $\lambda t \sim \omega_+^2$ as $t \to \infty$.}. Essentially, this approximation coincides with the rotating wave approximation from the quantum optics~\cite{Walls, Irish, Law}.

Note that the quadratic part of the interaction Hamiltonian, equation~\eqref{eq:qm-H-renorm}, can be absorbed into the free Hamiltonian, resulting in the renormalization of $\omega_+$, $\alpha$ and $\beta$:
\beq \label{eq:renorm}
\omega_+ \to \omega_+ + \frac{\lambda (N+2)}{4 N \omega_+^2} \left( |\alpha|^2 + |\beta|^2 \right), \quad \alpha \to \alpha + \frac{\lambda (N+2)}{8 N \omega_+^3} |\beta|^2 \alpha, \quad \beta \to \beta + \frac{\lambda (N+2)}{8 N \omega_+^3} |\alpha|^2 \beta. \eeq
Such a renormalization corresponds to the summation of ``daisy'' diagrams (compare with section~\ref{sec:SK} and section~4.5 of~\cite{Trunin-1}). We emphasize that in general frequency cannot be shifted independently from $\alpha$ and $\beta$ due to the strong backreaction caused by the low dimensionality of the problem. Also note that this renormalization assumes a relatively small coupling constant, i.e., $\lambda \ll \omega_+^3/|\beta|^2$. Since we work in the limit $\lambda \to 0$, $t \to \infty$, $\lambda t = \const$, this condition is ensured by large evolution times, $t \gg |\beta|^2/\omega_+$.

Discarding quadratic and constant terms, neglecting oscillating contributions and expanding~\eqref{eq:qm-Hint} to the second order in $\beta$, we derive an approximate Hamiltonian:
\beq \label{eq:H-approx}
\begin{aligned}
H_\text{int} &\approx \frac{\lambda}{8 N \omega_+^2} a_i^\dag a_i^\dag a_j a_j + \frac{\lambda}{4 N \omega_+^2} a_i^\dag a_j^\dag a_i a_j + \frac{3 \lambda \beta}{4 N \omega_+^2} a_i^\dag a_i a_j a_j + \frac{3 \lambda \beta^*}{4 N \omega_+^2} a_i^\dag a_i^\dag a_j^\dag a_j + \\ &+ \frac{3 \lambda |\beta|^2}{4 N \omega_+^2} a_i^\dag a_i^\dag a_j a_j + \frac{3 \lambda |\beta|^2}{2 N \omega_+^2} a_i^\dag a_j^\dag a_i a_j + \frac{3 \lambda \beta^2}{8 N \omega_+^2} a_i a_i a_j a_j + \frac{3 \lambda (\beta^*)^2}{8 N \omega_+^2} a_i^\dag a_i^\dag a_j^\dag a_j^\dag + \mO\left(|\beta|^3\right).
\end{aligned} \eeq
Keeping in mind the normal-ordered form of this Hamiltonian, we straightforwardly exponentiate it and obtain the evolved quantum state:
\beq \begin{aligned}
| \Psi(t) \rangle &= \mathcal{T} e^{-i \int_{t_0}^t H_\text{int}(t') dt'} | in \rangle \approx e^{- i t H_\text{int}} | in \rangle = \\ &= | in \rangle + 18 \frac{\beta^* |\beta|^2}{N} \left[ \exp\left(\frac{-i \lambda t}{4 \omega_+^2} \right) - 1 \right]^2 a_i^\dag a_i^\dag | in \rangle + \\ &\phantom{\approx | in \rangle} + \frac{3}{4} \frac{(\beta^*)^2}{N} \left[ \exp\left(\frac{-i \lambda t}{2 \omega_+^2} \right) - 1 \right] a_i^\dag a_i^\dag a_j^\dag a_j^\dag | in \rangle + \mO\left( |\beta|^4 \right) + \mO\left( \frac{1}{N^2} \right).
\end{aligned} \eeq
The leading contribution to this expression is ensured by the powers of the first term in the approximate Hamiltonian~\eqref{eq:H-approx}. The contribution of the second term is suppressed by the powers of $1/N$, and contribution of other terms is suppressed by the powers of $\beta$.

Substituting this expression into Eqs.~\eqref{eq:n} and~\eqref{eq:k}, we obtain the leading corrections to the tree-level level population and anomalous quantum average:
\begin{align}
\label{eq:n-small}
n_{ij}(t) &= \langle \Psi(t) | a_i^\dag a_j | \Psi(t) \rangle = \frac{\delta_{ij}}{N} \cdot 72 |\beta|^4 \sin^2\left( \frac{\lambda t}{4 \omega_+^2} \right) + \mO\left( |\beta|^6 \right) + \mO\left(\frac{1}{N^2}\right), \\
\label{eq:k-small}
\kappa_{ij}(t) &= \langle \Psi(t) | a_i a_j | \Psi(t) \rangle = \frac{\delta_{ij}}{N} \cdot 36 \beta^* |\beta|^2 \left[ \exp\left( \frac{-i \lambda t}{4 \omega_+^2} \right) - 1 \right]^2 + \mO\left( |\beta|^4 \right) + \mO\left(\frac{1}{N^2}\right).
\end{align}
Thus the total number of the created particles is given by the following expression:
\beq \label{eq:N-small}
\begin{aligned}
\mN &= \sum_{i = 1}^N \Big[ |\beta|^2 \delta_{ii} + \left(|\alpha|^2 + |\beta|^2 \right) n_{ii} + \alpha \beta \kappa_{ii} + \alpha^* \beta^* (\kappa_{ii})^* \Big] = \\ &= N |\beta|^2 + 36 |\beta|^4 \left[ 3 + \cos\left( \frac{\lambda t}{2 \omega_+^2} \right) - 4 \cos\left( \frac{\lambda t}{4 \omega_+^2} \right) \right] + \mO\left( |\beta|^5 \right) + \mO\left(\frac{1}{N}\right) \approx \\ &\approx N |\beta|^2 + 108 |\beta|^4 + \mO\left( |\beta|^5 \right) + \mO\left(\frac{1}{N}\right).
\end{aligned} \eeq
In the last line we replaced the oscillating contributions with their average values. Note that the correction to the tree-level particle number is always positive.

We emphasize that the calculations in this section are valid only for small deviations from stationarity, $|\beta| \ll 1$, where nondiagonal terms of the effective Hamiltonian are negligible. Unfortunately, this approximation does not cover the physically interesting resonant case~\eqref{eq:b-large}. Hence, we need to consider nondiagonal terms of the Hamiltonian and generalize identities~\eqref{eq:n-small},~\eqref{eq:k-small}, and~\eqref{eq:N-small} to arbitrary~$\beta$.

\section{Schwinger-Keldysh diagrammatic technique}
\label{sec:SK}

The Schwinger-Keldysh technique is a powerful tool to calculate correlation functions and quantum averages in nonstationary situations~\cite{Schwinger, Keldysh, Kamenev, Rammer, Landau:vol10, Arseev}. This technique can be concisely described by the following path integral~\cite{Berges, Leonidov, Radovskaya}:
\beq \label{eq:path}
\langle \hat{O} \rangle = \int \mathcal{D} \varphi(x) \mathcal{D} \pi(x) \mathcal{W}\left[ \varphi(x), \pi(x) \right] \int_{i.c.} \mathcal{D} \phi_\text{cl}(t,x) \mathcal{D} \phi_\text{q}(t,x) O e^{i S_K\left[ \phi_\text{cl}, \phi_\text{q} \right]}, \eeq
which calculates the expectation value of the operator $\hat{O}$ and explicitly contains the information about the initial state of the theory. Here $\varphi(x)$ and $\pi(x)$ denote the field and its conjugate momentum at the initial moment $t_0$; $\mathcal{W}\left[ \varphi(x), \pi(x) \right]$ denotes the Wigner function related to the initial value of the density matrix operator; $i.c.$ in the second integral means the initial conditions for the $\phi_\text{cl}$ field, $\phi_\text{cl}(t_0, x) = \varphi(x)$, $\dot{\phi}_\text{cl}(t_0,x) = \pi(x)$; and $S_K$ denotes the Keldysh action after the Keldysh rotation. For the theory~\eqref{eq:L} this action has the following form (there are no spatial directions in this case):
\beq S_K = -\int_{t_0}^\infty dt \left[ \phi_{i,\text{q}} \left( \pd_t^2 + \omega^2(t) \right) \phi_{i,\text{cl}} + \frac{\lambda}{N} \phi_{i,\text{cl}} \phi_{i,\text{cl}} \phi_{j,\text{cl}} \phi_{j,\text{q}} + \frac{\lambda}{4N} \phi_{i,\text{cl}} \phi_{i,\text{q}} \phi_{j,\text{q}} \phi_{j,\text{q}} \right] \eeq
Expanding the integrand of~\eqref{eq:path} as a series in $\lambda$ and assuming that the initial state is Gaussian, we straightforwardly obtain the Schwinger-Keldysh diagrammatic technique (Fig.~\ref{fig:technique}) with the following tree-level propagators:
\begin{figure}[t]
    \center{\includegraphics[scale=0.4]{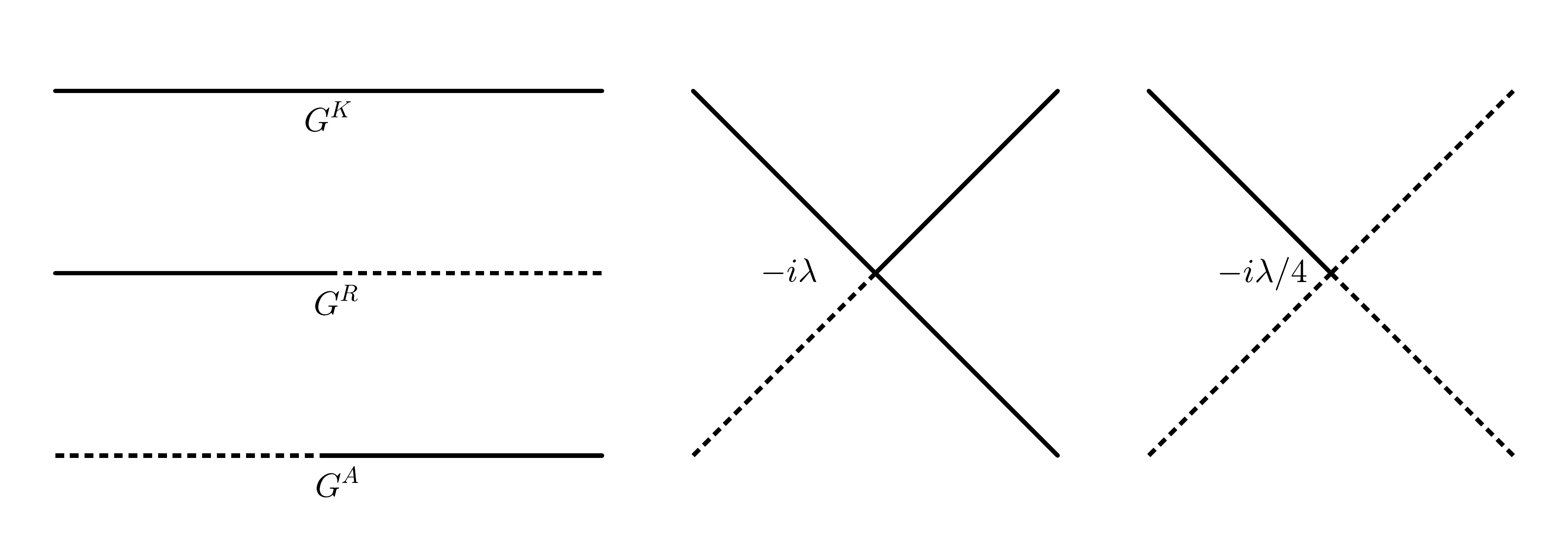}}
    \caption{Propagators and vertices in the Schwinger-Keldysh diagrammatic technique. The solid lines correspond to $\phi_\text{cl}$, the dashed lines correspond to $\phi_\text{q}$.}
    \label{fig:technique}
\end{figure}
\beq \begin{aligned}
i G_{0,ij}^K(t_1, t_2) &= \left\langle \phi_{i,\text{cl}}(t_1) \phi_{j,\text{cl}}(t_2) \right\rangle_0 = \frac{1}{2} \big\langle in \big| \big\{ \phi_i(t_1), \phi_j(t_2) \big\} \big| in \big\rangle, \\ 
i G_{0,ij}^R(t_1, t_2) &= \left\langle \phi_{i,\text{cl}}(t_1) \phi_{j,\text{q}}(t_2) \right\rangle_0 = \theta(t_1 - t_2) \big\langle in \big| \big[ \phi_i(t_1), \phi_j(t_2) \big] \big| in \big\rangle, \\
i G_{0,ij}^A(t_1, t_2) &= \left\langle \phi_{i,\text{q}}(t_1) \phi_{j,\text{cl}}(t_2) \right\rangle_0 = \theta(t_2 - t_1) \big\langle in \big| \big[ \phi_j(t_2), \phi_i(t_1) \big] \big| in \big\rangle,
\end{aligned} \eeq
where $\langle \cdots \rangle_0$ denotes the expectation value in the free theory and $\phi_i$ is the standard free quantized field~\eqref{eq:phi}. For simplicity, in the last identities we assumed that the initial state is pure.

These propagators have a simple physical meaning. On the one hand, retarded and advanced propagators describe the propagation of some localized perturbations (e.g., particles). Hence, at the tree level they do not depend on the state of the system. On the other hand, the Keldysh propagator explicitly contains the information about the state of the system:
\beq \label{eq:K-free}
i G_{0,ij}^K(t_1, t_2) = f(t_1) f^*(t_2) \left( \frac{1}{2} \delta_{ij} + n_{0,ij} \right) + f(t_1) f(t_2) \kappa_{0,ij} + H.c., \eeq
where $n_{0,ij} = \langle in | (a_i^\text{in})^\dag a_j^\text{in} | in \rangle$ and $\kappa_{0,ij} = \langle in | a_i^\text{in} a_j^\text{in} | in \rangle$ are initial level population and anomalous quantum average. If the initial state coincides with the ground state of the free Hamiltonian at the past infinity, $| in \rangle = | 0 \rangle$, at the tree level these quantum averages remain zero during the evolution of the system. However, at large evolution times they also receive substantial loop corrections. These corrections straightforwardly follow from the exact resummed Keldysh propagator because in the limit $t = \frac{t_1 + t_2}{2} \gg t_1 - t_2$ it has the form~\eqref{eq:K-free} with exact quantum averages $n_{ij}(t)$ and $\kappa_{ij}(t)$ defined in~\eqref{eq:n} and~\eqref{eq:k}.

Thus we need to resum leading loop corrections to the Keldysh propagator to calculate the exact quantum averages. We single out the leading loop contributions by considering the limit of small coupling constants, $\lambda \to 0$, $t \to \infty$, $\lambda t = \const$, and small time separations, $t_1 - t_2 \ll t$. Note that in this limit loop corrections to the retarded and advanced propagators are suppressed by the powers of~$\lambda$. We also assume the large~$N$ limit and keep the leading terms in the $1/N$ expansion.

There are two types of the $\mO(1)$ diagrams in the $O(N)$ model~\cite{Moshe, Polyakov:book, Coleman}. The first one --- the so-called ``daisy'' diagrams (Fig.~\ref{fig:daisy}) --- describe the leading corrections to the propagators. However, in these diagrams loop corrections are local and can be easily resummed with the following Dyson-Schwinger equations (the equation for the advanced propagator is similar to the equation for the retarded propagator):
\begin{figure}[t]
    \center{\includegraphics[scale=0.4]{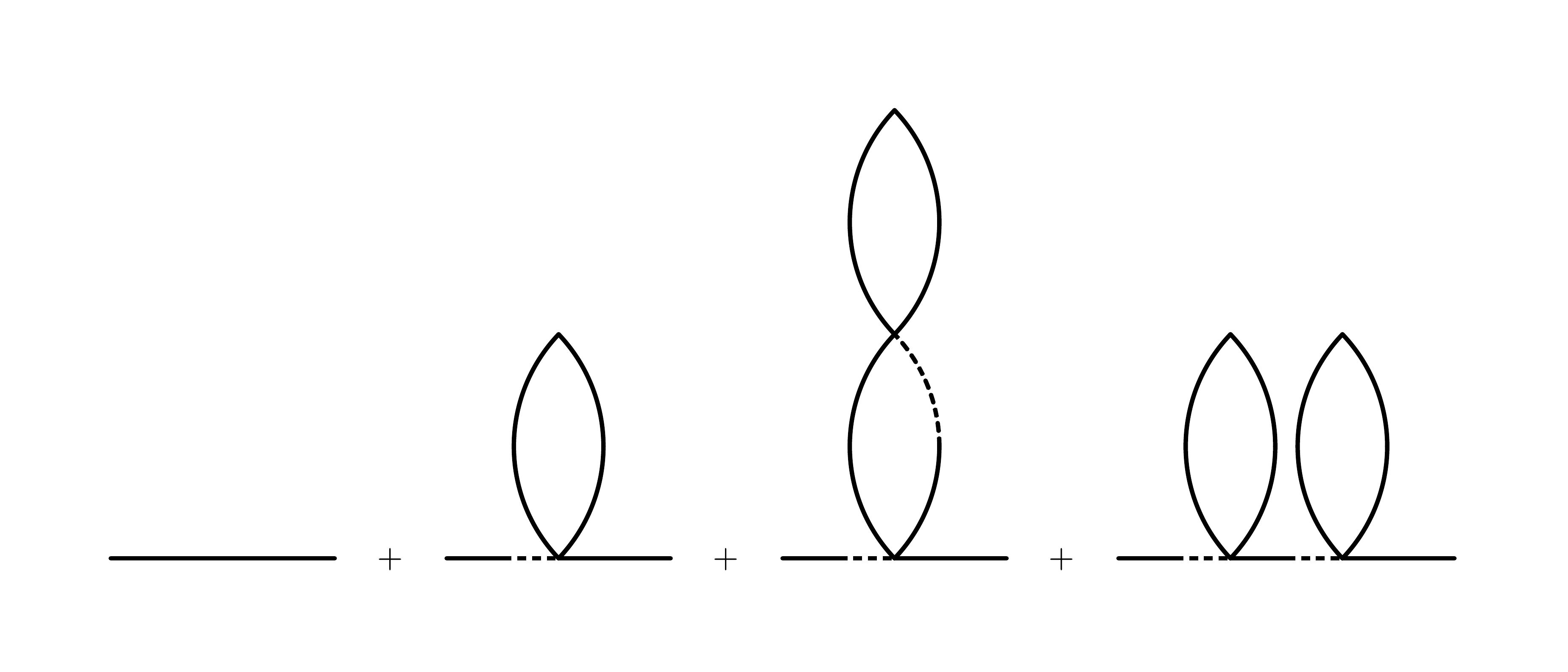}}
    \caption{Leading order, $\mathcal{O}(1)$, loop corrections to the Keldysh propagator. There are also conjugate diagrams for the Keldysh propagator and similar diagrams for the retarded and advanced propagators. Each internal line receives similar corrections.}
    \label{fig:daisy}
\end{figure}
\beq \begin{aligned}
\tilde{G}_{ij}^R(t_1, t_2) &= G_{0,ij}^R(t_1, t_2) - \frac{i \lambda}{N} \int_{t_0}^\infty dt G_{0,il}^R(t_1, t) \tilde{G}_{kk}^K(t,t) \tilde{G}_{lj}^R(t,t_2), \\
\tilde{G}_{ij}^K(t_1, t_2) &= G_{0,ij}^K(t_1, t_2) - \frac{i \lambda}{N} \int_{t_0}^\infty dt \left[ G_{0,il}^R(t_1, t) \tilde{G}_{kk}^K(t,t) \tilde{G}_{lj}^K(t,t_2) + G_{0,il}^K(t_1, t) \tilde{G}_{kk}^K(t,t) \tilde{G}_{lj}^A(t,t_2) \right].
\end{aligned} \eeq
Applying the operator $\pd_{t_1}^2 + \omega^2(t_1)$ to these equations, plugging $\tilde{G}^K(t,t) \equiv \tilde{G}_{kk}(t,t)/N$, and using the properties of the tree-level propagators, we straightforwardly obtain the following equations on the resummed propagators:
\beq \begin{aligned}
\left[ \pd_{t_1}^2 + \omega^2(t_1) \right] \tilde{G}_{ij}^R(t_1, t_2) &= -i \delta_{ij} \delta(t_1 - t_2) - \lambda \tilde{G}^K(t_1, t_1) \tilde{G}_{ij}^R(t_1, t_2), \\
\left[ \pd_{t_1}^2 + \omega^2(t_1) \right] \tilde{G}_{ij}^K(t_1, t_2) &= 0 - \lambda \tilde{G}^K(t_1, t_1) \tilde{G}_{ij}^K(t_1, t_2),
\end{aligned} \eeq
which imply a simple renormalization of the tree-level frequency:
\beq \omega^2(t) \to \omega^2(t) + \lambda \tilde{G}^K(t,t) \approx \omega_+^2 + \frac{\lambda}{2 \omega_+} \left( |\alpha|^2 + |\beta|^2 \right) + \mO(\lambda^2) + \mO\left(\frac{1}{N}\right), \quad \text{as} \quad t \to +\infty. \eeq
This result evidently reproduces the formula~\eqref{eq:renorm} in the leading order in $1/N$ and $\lambda$. We remind that the renormalization of the frequency also implies the renormalization of the coefficients~$\alpha$ and~$\beta$.

Practically this means that we can discard ``daisy'' diagrams if we consider the renormalized theory. This also means that we need to calculate the subleading, $\mO(1/N)$, correction to the Keldysh propagator to estimate the leading loop contributions to $n_{ij}$, $\kappa_{ij}$, and number of the created particles.

The other type of $\mO(1)$ diagrams are ``bubble'' diagrams that describe the corrections to the vertices (Fig.~\ref{fig:bubbles}). These diagrams are resummed with a similar Dyson-Schwinger equation:
\begin{figure}[t]
    \center{\includegraphics[scale=0.4]{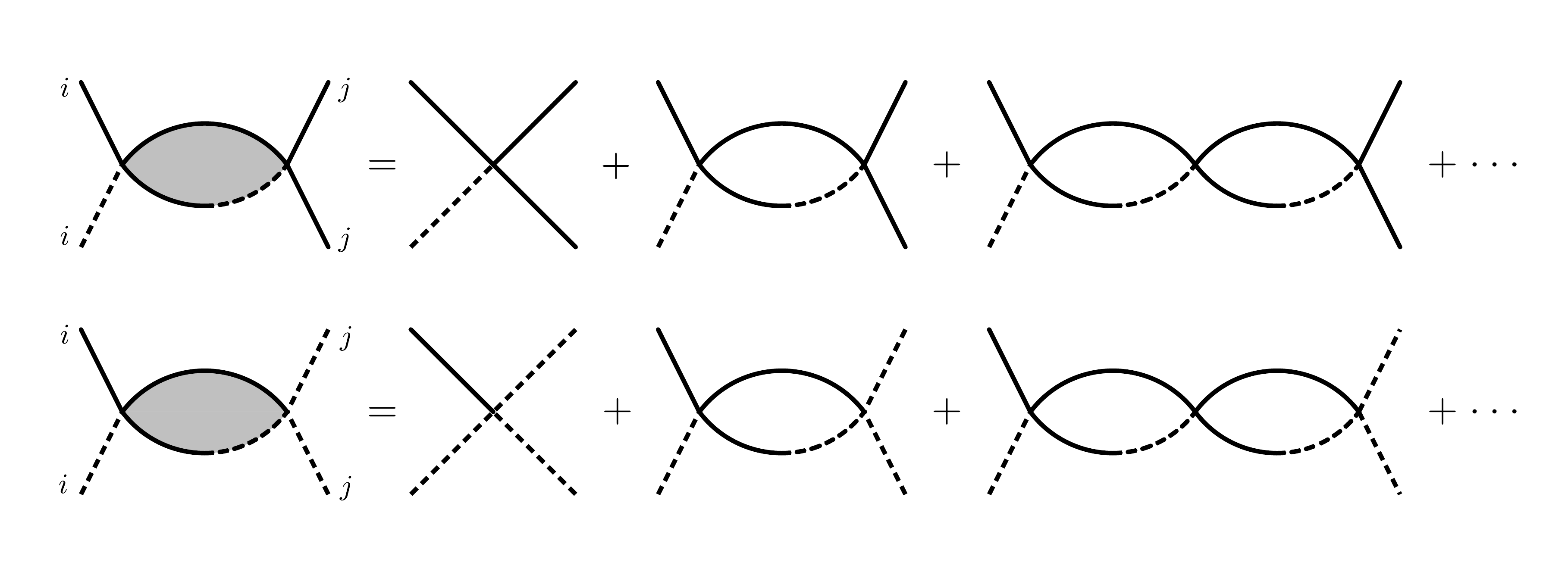}}
    \caption{Leading order, $\mathcal{O}(1)$, loop corrections to the vertices. Note that we included a ``zero-bubble'' diagram (a bare vertex) into the definition of the bubble chain, although in the Dyson-Schwinger equation~\eqref{eq:DS-bubble} we assumed that the decomposition starts from a single bubble. This is done to simplify Fig.~\ref{fig:diagrams}.}
    \label{fig:bubbles}
\end{figure}
\beq \label{eq:DS-bubble}
\tilde{B}(t_1, t_2) = 2 G_{0,kl}^R(t_1, t_2) G_{0,kl}^K(t_1, t_2) - \frac{2 i \lambda}{N} \int_{t_0}^\infty dt_3 G_{0,kl}^R(t_1, t_3) G_{0,kl}^K(t_1, t_3) \tilde{B}(t_3, t_2), \eeq
where $\tilde{B}(t_1, t_2)$ denotes the infinite chain of bubbles with truncated external legs. In Fig.~\ref{fig:bubbles}, this chain is denoted by the shaded loop. Note that equation~\eqref{eq:DS-bubble} contains combinatorial factors, which can be calculated by the method discussed in~\cite{Trunin-1}.

Equation~\eqref{eq:DS-bubble} is conveniently solved with the following ansatz inspired by the structure of the Keldysh and retarded propagators:
\beq \begin{aligned}
\tilde{B}(t_1, t_2) &= A \big(f^*(t_1)\big)^2 f^2(t_2) + B f^2(t_1) \big(f^*(t_2)\big)^2 + C f^2(t_1) f^2(t_2) + D \big(f^*(t_1)\big)^2 \big(f^*(t_2)\big)^2 = \\ &= \bem f^2(t_1) \\ \big(f^*(t_1)\big)^2 \eem^\dag \bem A(t_1, t_2) & D(t_1, t_2) \\ C(t_1, t_2) & B(t_1, t_2) \eem \bem f^2(t_2) \\ \big(f^*(t_2)\big)^2 \eem,
\end{aligned} \eeq
where $A$, $B$, $C$, and $D$ are some functions to be determined. In the second line we treated the mode functions as coordinates of a two vector to represent ansatz in a simple form. We remind that we consider the limit $t_1 - t_2 \ll t = \frac{t_1 + t_2}{2}$ and $\lambda \to 0$, $t \to \infty$, $\lambda t = \const$ to single out the leading loop contributions.

Substituting this ansatz into~\eqref{eq:DS-bubble}, keeping only non-oscillating terms in the integrand, discarding two vectors, and differentiating the identity over $t_1$, we obtain the following differential equation:
\beq \label{eq:generate-bubble}
\frac{d}{dt_1} \bem A & D \\ C & B \eem = \frac{\delta(t_1 - t_2)}{(2 \omega_+)^2} \bem 1 & 0 \\ 0 & -1 \eem - \frac{i \lambda}{(2 \omega_+)^2} \bem 1 + 6 |\beta|^2 + 6 |\beta|^4 & 6 \alpha^2 \beta^2 \\ -6 (\alpha^*)^2 (\beta^*)^2 &  -1 -6 |\beta|^2 - 6 |\beta|^4 \eem \bem A & D \\ C & B \eem. \eeq
The solution to this equation is given by the matrix exponential:
\beq \label{eq:exact-bubble}
\begin{aligned}
\bem A & D \\ C & B \eem &= \frac{\theta(t_{12})}{(2 \omega_+)^2} \exp \left[ - \frac{i \lambda t_{12}}{(2 \omega_+)^2} \bem 1 + 6 |\beta|^2 + 6 |\beta|^4 & 6 \alpha^2 \beta^2 \\ -6 (\alpha^*)^2 (\beta^*)^2 & -1 -6 |\beta|^2 - 6 |\beta|^4 \eem \right] \bem 1 & 0 \\ 0 & -1 \eem = \\ &= \frac{\theta(t_{12})}{(2 \omega_+)^2} \bem \cos \frac{\lambda R t_{12}}{4 \omega_+^2} - i \frac{1 + 6 |\beta|^2 + 6 |\beta|^4}{R} \sin \frac{\lambda R t_{12}}{4 \omega_+^2} & i \frac{6 \alpha^2 \beta^2}{R} \sin \frac{\lambda R t_{12}}{4 \omega_+^2} \\ i \frac{6 (\alpha^*)^2 (\beta^*)^2}{R} \sin \frac{\lambda R t_{12}}{4 \omega_+^2} & -\cos \frac{\lambda R t_{12}}{4 \omega_+^2} - i \frac{1 + 6 |\beta|^2 + 6 |\beta|^4}{R} \sin \frac{\lambda R t_{12}}{4 \omega_+^2}  \eem,
\end{aligned} \eeq
where we introduce the short notation for the time difference, $t_{12} \equiv t_1 - t_2$, and positive eigenvalue of the generating matrix:
\beq R \equiv \sqrt{1 + 12 |\beta|^2 + 12 |\beta|^4}. \eeq
Note that for small deviations from stationarity, $|\beta| \ll 1$, the generating matrix is approximately diagonal. This is because in this limit the leading contributions to the ``bubble'' diagrams are associated with the diagonal part of the effective Hamiltonian~\eqref{eq:qm-Hint}.

Finally, let us estimate the subleading, $\mO(1/N)$, contribution to the Keldysh propagator. Substituting the resummed ``bubble'' diagrams into the two-loop corrections to the Keldysh propagator (Fig.~\ref{fig:diagrams}) and performing some tedious but straightforward calculations, we obtain the exact resummed Keldysh propagator:
\begin{figure}[t]
    \center{\includegraphics[scale=0.4]{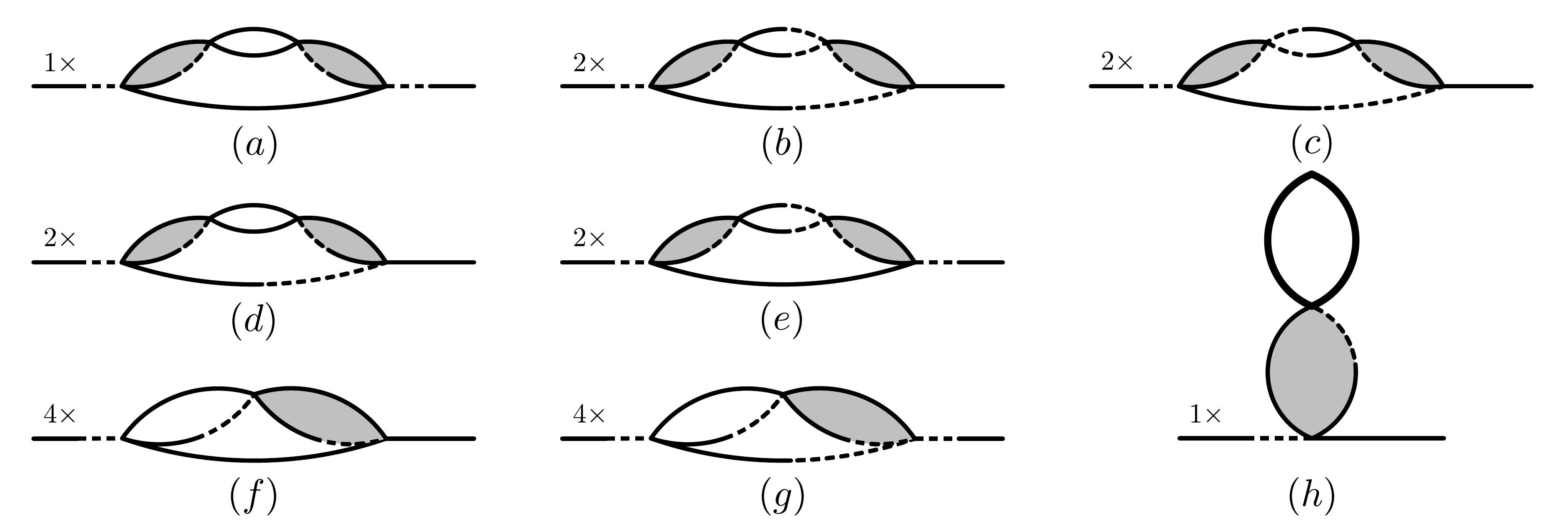}}
    \caption{Subleading order, $\mathcal{O}(1/N)$, loop corrections to the Keldysh propagator with combinatorial factors. The shaded loops correspond to the resummed ``bubble'' diagrams from Fig.~\ref{fig:bubbles}. The bold solid line in (h) denotes the sum of (a)-(g) corrections. Note that each of the listed diagrams has a conjugate counterpart.}
    \label{fig:diagrams}
\end{figure}
\beq \label{eq:K-exact}
i G_{ij}^K(t_1, t_2) = f(t_1) f^*(t_2) \left( \frac{1}{2} \delta_{ij} + n_{ij} \right) + f(t_1) f(t_2) \kappa_{ij} + H.c., \eeq
with the following quantum averages:
\begin{align}
\label{eq:n-large}
&n_{ij} = \frac{\delta_{ij}}{N} \cdot 72 \frac{|\alpha|^4 |\beta|^4}{R^2} \sin^2 \left( \frac{\lambda t}{4 \omega_+^2} R \right) + \mO\left(\frac{1}{N}\right), \\
\label{eq:k-large}
&\begin{aligned}
\kappa_{ij} = \frac{\delta_{ij}}{N} \cdot 36 \frac{\alpha^* \beta^* |\alpha|^2 |\beta|^2 \left( |\alpha|^2 + |\beta|^2 \right)}{R^2} \bigg[ \frac{1 + 6 |\beta|^2 + 6 |\beta|^4}{R^2} \cos\left( \frac{\lambda t}{2 \omega_+^2} R \right) - \frac{i}{R} \sin\left( \frac{\lambda t}{2 \omega_+^2} \right) - \\ - \frac{2}{R^2} \cos\left( \frac{\lambda t}{4 \omega_+^2} R \right) + \frac{2 i}{R} \sin\left( \frac{\lambda t}{4 \omega_+^2} R \right) + \frac{1 - 6 |\beta|^2 - 6 |\beta|^4}{R^2} \bigg] + \mO\left(\frac{1}{N}\right). \end{aligned}
\end{align}
We emphasize that both quantities are finite even at very large evolution times, although each term of the series in $\lambda$ grows secularly and goes to infinity as $t \to \infty$. Besides that, both quantities are suppressed by $1/N$ in comparison with the tree-level contribution to~\eqref{eq:K-exact}.

At small $\beta$ identities~\eqref{eq:n-large} and~\eqref{eq:k-large} evidently reproduce the expressions~\eqref{eq:n-small} and~\eqref{eq:k-small} obtained in the previous section. They also extend these results to large $\beta$. At first glance, for large~$\beta$ leading contributions to $n_{ij}$ and $\kappa_{ij}$ dominate; e.g., in the resonant case~\eqref{eq:b-large} they are exponentially amplified:
\beq \label{eq:nk-leading}
\begin{aligned}
&n_{ij} = \frac{\delta_{ij}}{N} \cdot \frac{3}{4} e^{4 \gamma \omega_+ t_R} \sin^2 \left( \frac{\lambda t \sqrt{3}}{8 \omega_+^2} e^{2 \gamma \omega_+ t_R} \right) + \mO\left( e^{2 \gamma \omega_+ t_R} \right) + \mO\left(\frac{1}{N}\right), \\
&\kappa_{ij} = \frac{\delta_{ij}}{N} \cdot \frac{3 i}{4} e^{4 \gamma \omega_+ t_R} \sin^2 \left( \frac{\lambda t \sqrt{3}}{8 \omega_+^2} e^{2 \gamma \omega_+ t_R} \right) + \mO\left( e^{2 \gamma \omega_+ t_R} \right) + \mO\left(\frac{1}{N}\right),
\end{aligned} \eeq
and surpass the tree-level expression if the resonant oscillations continue for a long enough time, $t_R \gg \frac{\log N}{4 \gamma \omega_+}$. Nevertheless, the leading contributions~\eqref{eq:nk-leading} cancel each other in the resummed Keldysh propagator with explicitly expanded modes:
\beq \label{eq:K-large}
\begin{aligned}
i G_{ij}^K(t_1, t_2) &= \left[ \frac{1}{2} + \frac{1}{N} \frac{288 |\alpha|^4 |\beta|^4}{R^4} \sin^4\left( \frac{\lambda t}{8 \omega_+^2} R \right) \right] \big( |\alpha|^2 + |\beta|^2 \big) \delta_{ij} \frac{e^{-i \omega_+ (t_1 - t_2)}}{2 \omega_+} + \\ &+ \bigg[ 1 + \frac{1}{N} \frac{36 |\alpha|^2 |\beta|^2}{R^3} \bigg( \frac{1}{R} + \frac{1 + 8 |\alpha|^2 |\beta|^2}{R} \cos\left( \frac{\lambda t}{2 \omega_+^2} R \right) - i \big( |\alpha|^2 + |\beta|^2 \big) \sin\left( \frac{\lambda t}{2 \omega_+^2} R \right) - \\ &- 2 \frac{\big( |\alpha|^2 + |\beta|^2 \big)^2}{R} \cos\left( \frac{\lambda t}{4 \omega_+^2} R \right) + 2 i  \big( |\alpha|^2 + |\beta|^2 \big) \sin\left( \frac{\lambda t}{4 \omega_+^2} R \right) \bigg) \bigg] \alpha \beta^* \delta_{ij} \frac{e^{-i \omega_+ (t_1 + t_2)}}{2 \omega_+} + \\ &+ H.c. + \mO\left( \frac{1}{N} \right),
\end{aligned} \eeq
and resummed number of the created particles:
\beq \label{eq:N-large}
\mN = N |\beta|^2 + 36 \frac{|\alpha|^4 |\beta|^4 \left( |\alpha|^2 + |\beta|^2 \right)}{R^4} \left[ 3 + \cos\left( \frac{\lambda t}{2 \omega_+^2} R \right) - 4 \cos\left( \frac{\lambda t}{4 \omega_+^2} R \right) \right] + \mO\left( \frac{1}{N} \right). \eeq
This cancellation resembles the cancellation of the leading two-loop corrections to the Keldysh propagator of light scalar fields in de~Sitter space~\cite{Popov, Pavlenko}.

Note that level density and anomalous quantum average in the model~\eqref{eq:L} cannot be measured directly; instead of it, these averages should be extracted from explicitly observable quantities, such as the average number (or energy) of the created particles. These quantities, in turn, are built from the derivatives of the resummed Keldysh propagator that is naturally defined through the in-modes~\eqref{eq:qm-modes}. At the past infinity in-mode looks like a single plane wave, but at the future infinity it contains both positive- and negative-frequency solutions. Expressions~\eqref{eq:n-large}, \eqref{eq:k-large} do not have this difference. At the same time, observables are conveniently calculated if we expand in-modes in the Keldysh propagator and rearrange parts proportional to single exponents at the future infinity. This allows one to discard rapidly oscillating terms that are negligible at large evolution times. As was shown in~\eqref{eq:K-large}, after such a rearrangement growing with $|\beta|$ (proportional to $|\beta|^a$, $a > 0$) contributions to $n_{ij}$ and $\kappa_{ij}$ cancel each other. Therefore, the apparent exponential amplification~\eqref{eq:nk-leading} is nonphysical.

Thus, in strongly nonstationary case, $|\beta| \gg 1$, the result of the loop corrections is proportional to the same power of $\beta$ as the tree-level contribution, but suppressed by $1/N$:
\beq \begin{aligned}
\mN(t) &= N |\beta|^2 + \frac{1}{2} |\beta|^2 \left[ 3 + \cos\left( \frac{\lambda t}{\omega_+^2} |\beta|^2 \sqrt{3} \right) - 4 \cos\left( \frac{\lambda t}{2 \omega_+^2} |\beta|^2 \sqrt{3} \right) \right] + \mO\left( |\beta|^0 \right) + \mO\left( \frac{1}{N} \right) \approx \\ &\approx N |\beta|^2 + \frac{3}{2} |\beta|^2 + \mO\left( |\beta|^0 \right) + \mO\left( \frac{1}{N} \right).
\end{aligned} \eeq
For the last identity we neglected oscillating expressions.

We emphasize that in the resonant case~\eqref{eq:b-large} both tree-level and loop-level contributions to the Keldysh propagator and number of created particles indefinitely grow with the duration of oscillations $t_R$. The primary source of this growth is the mixing of positive- and negative-frequency modes, which is characterized by the Bogoliubov coefficient $\beta$ (we note that in the resonant case $\alpha \sim \beta \sim e^{\omega \gamma t_R}$). This growth is related to the variations of frequency and was presented already at the tree-level. Additionally, interactions between the modes generate finite, but nonzero $n_{ij}$ and $\kappa_{ij}$. As was previously discussed, the leading physically meaningful contribution to these quantities is proportional to $|\beta|^0$. Substituting this contribution into~\eqref{eq:N-free} and~\eqref{eq:N-exact}, we see that both $\mN^\text{free}$ and $\mN$ exponentially grow with $t_R$. Loop corrections modify the prefactor of exponential growth, but do not alter the qualitative behavior of $\mN$. We also note that $t_R$ should not be confused with the evolution time $t$, which is much larger than $t_R$.

Roughly speaking, loop corrections act as $\mO(1)$ additional degrees of freedom, $N \to N + \frac{3}{2}$, during the measurement of particle number well after the end of resonant oscillations. Of course, the average contribution of these ``phantom'' degrees of freedom is also accompanied by harmonic oscillations of a comparable amplitude. Nevertheless, the frequency of these oscillations rapidly grow with~$\beta$ (exponentially in the resonant case), so they will be difficult to detect in feasible experiments.

\section{Discussion and Conclusion}
\label{sec:discussion}

In this paper we calculated exact quantum averages in a simple example of nonstationary large~$N$ model --- namely, a quantum anharmonic oscillator with a quartic $O(N)$ interaction term and time-dependent frequency. We performed this calculation using two different methods. First, we deduced an effective Hamiltonian of the model using a kind of rotating wave approximation. For small deviations from the stationarity this Hamiltonian is approximately diagonal, so the quantum averages straightforwardly follow from the decomposition of the evolution operator. Second, we reproduced these results and extended them to arbitrarily large deviations from the stationarity using the Schwinger-Keldysh diagrammatic technique.

In both cases we resummed the leading loop contributions to the exact $n_{ij}$ and $\kappa_{ij}$. We demonstrated that for large evolution times these quantities oscillate near nonzero average values. In the case of resonantly oscillating frequency average values exponentially grow and surpass the tree-level expressions even in the large~$N$ limit. However, these exponentially growing contributions cancel each other in the exact Keldysh propagator~\eqref{eq:K-large} and particle number~\eqref{eq:N-large}. Thus, in strongly nonstationary situations loop contributions to $G_{ij}^K$ and $\mN$ are proportional to the same power of~$\beta$ as $G_{0,ij}^K$. Roughly speaking, these contributions result in $\mO(1)$ additional degrees of freedom and increase the rate at which the system absorbs energy from the external world. In weakly nonstationary situations loop contributions are additionally suppressed by the powers of~$\beta$.

We emphasize that the system considered in this paper is explicitly nonstationary: not only its initial state is out-of-equilibrium, but the effective mass of the field varies with time. In other words, we considered an open system that can exchange energy with the external world. On the one hand, this evident nonstationarity distinguishes our model from the previous work on nonequilibrium large~$N$ models~\cite{Cooper:1994, Cooper:1996, Baacke:1997, Baacke:2000, Boyanovsky, Das}, which was mainly devoted to closed systems. On the other hand, the Hamiltonian of our toy model is similar to the Hamiltonians of interacting quantum fields in nonstationary spacetimes (e.g., expanding universe, collapsing star, or moving mirror). Due to this reason, we believe that our analysis provides useful insights into the physics of these complex systems.

The analysis of this paper can be extended in several possible directions. First, it is promising to generalize our results to finite $N$ nonstationary systems, e.g., explicitly calculate $\mN$ in the $N=1$ version of the model~\eqref{eq:L}. For small deviations from the stationarity, i.e., $\beta \ll 1$, this extension is obvious because the analysis of section~\ref{sec:H} does not essentially require a $1/N$ expansion. An example of such calculation is presented in appendix~\ref{sec:small-b}. At the same time, the diagrammatic calculations of section~\ref{sec:SK} heavily rely on the~$1/N$ expansion that singles out a particular set of diagrams (Fig.~\ref{fig:bubbles} for the vertices and Fig.~\ref{fig:diagrams} for the Keldysh propagator) and decouples the system of Dyson-Schwinger equations. In the finite~$N$ case the solution to this system is unknown. So, the extension of the effective Hamiltonian analysis to large $\beta$ is unclear.

Note that in the large~$N$ limit particle decays ($a^\dag a a a$ and $a^\dag a^\dag a^\dag a$ terms of the effective Hamiltonian) are suppressed in favor of scattering ($a^\dag a^\dag a a$ term) and creation of virtual particles ($a a a a$ and $a^\dag a^\dag a^\dag a^\dag$ terms). At the same time, in the finite~$N$ models all these processes equally contribute to the resummed Keldysh propagator. However, we believe that this simplification is not very important because ``bubble'' diagrams, which describe scattering and creation of virtual particles, are multiplied by large combinatorial factors\footnote{These factors can be calculated similarly to~\cite{Trunin-1}.}. For instance, in the $N=1$ version of the model~\eqref{eq:L}, a chain of $2m$ ``bubbles'' is multiplied by $36^m$. In comparison, a chain of $m$ ``sunset'' diagrams, which has the same order in $\lambda$, is multiplied by $18^m$. Due to this reason we expect that at large orders of perturbation theory bubbles become parametrically larger than other diagrams even in the finite~$N$ case. So we believe that qualitative behavior of the Keldysh propagator and created particle number coincide in the large~$N$ and finite~$N$ models. The analysis of appendix~\ref{sec:small-b} supports this reasoning for small $\beta$.

Second, it is interesting to extend our analysis to nonvacuum initial states, including finite-temperature thermal states and mixed states with nontrivial initial $n_{0,ij}$ and $\kappa_{0,ij}$. This extension should be performed carefully because in zero-dimensional quantum systems a naive version of the Wick theorem does not work~\cite{Trunin-1}. However, it is still applicable after some modifications~\cite{Arseev}.

Finally, the approach of section~\ref{sec:H} can be applied to higher-dimensional systems, such as light scalar fields in de~Sitter space~\cite{Popov, Pavlenko, Moschella, Serreau-1, Serreau-2, Serreau-3} and the dynamical Casimir effect~\cite{Alexeev, Akopyan}. We expect that for small deviations from stationarity these models also allow one to exponentiate the effective Hamiltonian and obtain analogs of identities~\eqref{eq:n-small} and~\eqref{eq:k-small}. However, diagrammatic calculations in these models face with severe difficulties even in the large~$N$ limit. Indeed, in higher-dimensional models mode functions carry additional momentum indices, so the generating matrix of ``bubble'' diagrams~\eqref{eq:generate-bubble} turns into a tensor with $2+4D$ indices, where $D$ is spatial dimensionality. Thus, ``bubble'' summation requires an exponentiation of nondiagonal tensor. We believe that in some particular cases this tensor can be simplified, so the exponentiation is feasible. Nevertheless, in general, such a simplification is not guaranteed.

Besides that, we believe that the results of this paper, especially identities~\eqref{eq:K-large} and~\eqref{eq:N-large}, have experimental implications. As we have previously mentioned, in the resonant case loop corrections result in additional ``phantom'' degrees of freedom, $N \to N + \frac{3}{2}$, which modify the average number, $\mN = N |\beta|^2 \to \left(N + \frac{3}{2}\right) |\beta|^2$, and average energy, $E = \omega_+ \mN$, of the created particles. Although in the large~$N$ limit this contribution is suppressed, in the finite~$N$ case it is distinguishable. Due to this reason, we expect that it can be measured in practice. Finally, we remind that this result was obtained in the limit of large $N$, small coupling constant, $\lambda \ll \frac{|\beta|^2 \omega_+^3}{1 + 6 |\beta|^2 + 6 |\beta|^4}$, and large evolution time, $\omega_+^2/\lambda \ll t \ll \omega_+^5/\lambda^2$.

\section*{Acknowledgments}

We would like to thank Emil Akhmedov and Sebasti\'an Franchino-Vi\~nas for the fruitful discussions. This work was supported by the Russian Ministry of Education and Science and by the grant from the Foundation for the Advancement of Theoretical Physics and Mathematics ``BASIS''.

\appendix

\section{Expectation value of the full Hamiltonian at the future infinity}
\label{sec:N}

In this section we use the decomposition of the exact Keldysh propagator:
\beq i G_{nk}^K(t_1, t_2) \approx \left[ \left(\frac{1}{2} \delta_{nk} + n_{nk}(t) \right) f^\text{in}_n(t_1) \left( f_k^\text{in}(t_2) \right)^* + \kappa_{nk}(t) f_n^\text{in}(t_1) f_k^\text{in}(t_2) + H.c. \right], \eeq
where $t = \frac{t_1 + t_2}{2}$ and $n_{nk}(t)$, $\kappa_{nk}(t)$ are given by identities~\eqref{eq:n},~\eqref{eq:k}, to estimate the expectation value of the evolved free Hamiltonian, $\bar{H} = \langle in | U^\dag(t,t_0) H_\text{free} U(t,t_0) | in \rangle$, in the limit $t \to +\infty$. For simplicity we consider the quantum mechanical theory~\eqref{eq:L}, whose Hamiltonian is represented as follows:
\beq \begin{aligned}
\bar{H}(x) &= \frac{1}{2} \pd_{t_1} \pd_{t_2} 
\sum_{n=1}^N i G_{nn}^K(t_1, t_2) \Big|_{t_1 = t_2 = t} + \frac{\omega^2(t)}{2} \sum_{n=1}^N i G_{nn}^K(t, t)= \\
&= \frac{1}{2} \sum_{n,k=1}^N \left[ \left(\frac{1}{2} \delta_{nk} + n_{nk} \right) \dot{f}^\text{in}(t) \left( \dot{f}^\text{in}(t) \right)^* + \kappa_{nk} \dot{f}^\text{in}(t) \dot{f}^\text{in}(t) + H.c. \right] + \\ &+ \frac{\omega^2(t)}{2} \sum_{n,k=1}^N \left[ \left(\frac{1}{2} \delta_{nk} + n_{nk} \right) f^\text{in}(t) \left( f^\text{in}(t) \right)^* + \kappa_{nk} f^\text{in}(t) f^\text{in}(t) + H.c. \right].
\end{aligned} \eeq
Substituting the future asymptotic of the in-modes~\eqref{eq:qm-modes} into this expression, we obtain the following Hamiltonian:
\beq \label{eq:H-full}
\begin{aligned}
\bar{H}(x) &= \sum_{n,k} \omega_+ \left[ \left( \frac{1}{2} \delta_{nk} + n_{nk} \right) \left( |\alpha|^2 + |\beta|^2 \right) + \alpha \beta \kappa_{nk} + \alpha^* \beta^* \kappa_{nl}^* \right] = \\ &= \frac{1}{2} \omega_+ N + \omega_+ \mN,
\end{aligned}\eeq
where $\mN = \sum_{n=1}^N \mN_n$, $\mN_n$ is defined by~\eqref{eq:N-exact}, and we use the property of the Bogoliubov coefficients, $|\alpha|^2 - |\beta|^2 = 1$. Note that the term $\frac{1}{2} \omega_+ N$ corresponds to the energy of the zero-point fluctuations.

Finally, recall that the full Hamiltonian also contains the quartic interaction term:
\beq \bar{H}_\text{full} = \bar{H} + \bar{H}_\text{int}, \quad \bar{H}_\text{int}(t) \equiv \frac{\lambda}{4 N} \langle in | \phi_i(t) \phi_i(t) \phi_j(t) \phi_j(t) | in \rangle. \eeq
In the limit $t \to +\infty$ this term is also proportional to $N$:
\beq \bar{H}_\text{int} = \frac{\lambda}{4 N} i G_{ii}^K(t,t) i G_{jj}^K(t,t) + \mO\left(\frac{1}{N}\right) \approx \frac{\lambda N}{16 \omega_+^2} \left( |\alpha|^4 + 4 |\alpha|^2 |\beta|^2 + |\beta|^4 \right) + \mO\left(1\right). \eeq
Furthermore, for large $\beta$ it is proportional to $|\beta|^4$ and seemingly exceeds the contribution of $\mN$ to the full Hamiltonian. However, we remind that we work in the limit $\lambda \to 0$, $t \to \infty$, $\lambda t = \const$. Hence, at large evolution times, $t \gg \frac{1 + 6 |\beta|^2 + 6 |\beta|^4}{|\beta|^2 \omega_+}$, the contribution of $H_\text{int}$ is suppressed by a small coupling constant, $\lambda \ll \frac{|\beta|^2 \omega_+^3}{1 + 6 |\beta|^2 + 6 |\beta|^4}$, and can be neglected. Note that at large $\beta$ this time scale coincides with the one established for the renormalization relations~\eqref{eq:renorm}. At the same time, for small $\beta$ this condition is more restrictive than requirement for~\eqref{eq:renorm}.

Thus in the limit in question identity~\eqref{eq:H-full} provides an approximate expression for the full Hamiltonian at the future infinity. This confirms that under the mentioned assumptions $\mN$ can be interpreted as the total number of created out-particles (i.e., excitations over the vacuum) in the full interacting theory. We also believe that the reasoning in this section can be extended to higher-dimensional quantum field theories with nondegenerate spectrum.

\section{Single oscillator in the limit of weak nonstationarity}
\label{sec:small-b}

In this section we repeat the method of section~\ref{sec:H} for an $N=1$ version of the model~\eqref{eq:L}:
\beq \mathcal{L} = \frac{1}{2} \dot{\phi}^2 - \frac{\omega^2(t)}{2} \phi^2 - \frac{\lambda}{4} \phi^4, \eeq
where $\omega(t) \to \omega_\pm$ as $t \to \pm \infty$. Substituting the quantized field (note that free modes coincide with those of the large~$N$ model) into the Hamiltonian and taking the limit $\lambda \to 0$, $t \to \infty$, $\lambda t = \const$, we obtain the following expressions for the interacting Hamiltonian:
\beq \label{eq:H-finite-N}
\begin{aligned}
H_\text{int} &\approx \frac{3 \lambda \left( |\alpha|^4 + 4 |\alpha|^2 |\beta|^2 + |\beta|^4 \right)}{16 \omega_+^2} \big(a^\dag\big)^2 a^2 + \frac{3 \lambda \alpha \beta \left( |\alpha|^2 + |\beta|^2 \right)}{4 \omega_+^2} a^\dag a^3 + \frac{3 \lambda \alpha^2 \beta^2}{8 \omega_+^2} a^4 + H.c. \approx \\ &\approx \frac{3 \lambda}{16 \omega_+^2} \big(a^\dag\big)^2 a^2 + \frac{3 \lambda \beta}{4 \omega_+^2} a^\dag a^3 + \frac{9 \lambda |\beta|^2}{8 \omega_+^2} \big(a^\dag\big)^2 a^2 + \frac{3 \lambda \beta^2}{8 \omega_+^2} a^4 + \mO\left( |\beta|^3 \right) + H.c.,
\end{aligned} \eeq
and renormalized parameters of the free theory:
\beq \omega_+ \to \omega_+ + \frac{3 \lambda}{4 \omega_+^2} \left( |\alpha|^2 + |\beta|^2 \right), \quad \alpha \to \alpha + \frac{3 \lambda}{8 \omega_+^3} |\beta|^2 \alpha, \quad \beta \to \beta + \frac{3 \lambda}{8 \omega_+^3} |\alpha|^2 \beta. \eeq
Exponentiating the Hamiltonian~\eqref{eq:H-finite-N} and keeping only the leading powers of $\beta$, we obtain an approximate expression for the evolved quantum state:
\beq \begin{aligned}
| \Psi(t) \rangle &\approx | in \rangle + 2 \beta^* |\beta|^2 \left[ \frac{1}{5} \exp\left(\frac{-9 i \lambda t}{2 \omega_+^2} \right) - \frac{6}{5} \exp\left(\frac{-3 i \lambda t}{4 \omega_+^2} \right) + 1 \right] \big( a^\dag \big)^2 | in \rangle + \\ &\phantom{\approx | in \rangle} + \frac{(\beta^*)^2}{12} \left[ \exp\left(\frac{-9 i \lambda t}{2 \omega_+^2} \right) - 1 \right] \big( a^\dag \big)^4 | in \rangle + \mO\left( |\beta|^4 \right).
\end{aligned} \eeq
This expression straightforwardly implies the exact level population, anomalous quantum average:
\begin{align}
n(t) &= \langle \Psi(t) | a^\dag a | \Psi(t) \rangle \approx \frac{8}{3} |\beta|^4 \sin^2\left( \frac{9 \lambda t}{4 \omega_+^2} \right) + \mO\left( |\beta|^6 \right), \\
\kappa(t) &= \langle \Psi(t) | a a | \Psi(t) \rangle \approx 4 \beta^* |\beta|^2 \left[ \frac{1}{5} \exp\left(\frac{-9 i \lambda t}{2 \omega_+^2} \right) - \frac{6}{5} \exp\left(\frac{-3 i \lambda t}{4 \omega_+^2} \right) + 1 \right] + \mO\left( |\beta|^4 \right),
\end{align}
and exact number of the created particles:
\beq \begin{aligned}
\mN &= |\beta|^2 + \big( |\alpha|^2 + |\beta|^2 \big) n + \alpha \beta \kappa + \alpha^* \beta^* \kappa^* = \\ &= |\beta|^2 + |\beta|^4 \left[ \frac{28}{3} + \frac{4}{15} \cos\left(\frac{9 \lambda t}{2 \omega_+^2} \right) - \frac{48}{5} \cos\left(\frac{3 \lambda t}{4 \omega_+^2} \right) \right] + \mO\left( |\beta|^5 \right).
\end{aligned} \eeq
We emphasize that the correction to the tree-level particle number is always positive. This result is very similar to the identity~\eqref{eq:N-small} from the large~$N$ version of the model, although the average values of the corrections and forms of the oscillating curves in these models are different.

\end{document}